\documentclass[11pt,a4paper]{article}
\pdfoutput=1

\usepackage{jcappub}
\usepackage{multirow}
\usepackage{amssymb}

\newcommand{\fett}[1]{\boldsymbol{#1}}

\newcommand{\dd}{{\rm{d}}}
\newcommand{\ii}{{\rm{i}}}

\newcommand{\be}{\begin{equation}}
\newcommand{\ee}{\end{equation}}

\newcommand{\kop}
{\mathfrak{K}}

\newcommand{\Hc}{\mathcal{H}}

\newcommand{\CLASS}{{\sc class}}

\newcommand{\inspire}[1]{\href{http://inspirehep.net/search?p=find+J+#1}
 {{\color{black}[{\color{blue} {\small in}SPIRE}]}}}
\newcommand{\book}[1]{\href{http://inspirehep.net/search?p=#1}
 {{\color{black}[{\color{blue} {\small in}SPIRE}]}}}

\newcommand{\inspired}[1]{\href{http://inspirehep.net/search?p=#1}
 {{\color{black}[{\color{blue} {\small in}SPIRE}]}}}

\newcommand{\HL}{H_{\rm L}}
\newcommand{\HT}{H_{\rm T}}

\newcommand{\nab}{\nabla}

\begin{document}

\title{General relativistic weak-field limit and Newtonian N-body simulations}

\date{\today}

\author[a]{Christian Fidler,}
\emailAdd{christian.fidler@uclouvain.be}

\author[e]{Thomas Tram,}
\emailAdd{thomas.tram@phys.au.dk}

\author[c,d\qquad\qquad]{Cornelius Rampf,}
\emailAdd{rampf@thphys.uni-heidelberg.de} 

\author[b]{Robert Crittenden,}
\emailAdd{robert.crittenden@port.ac.uk}

\author[b]{Kazuya Koyama}
\emailAdd{kazuya.koyama@port.ac.uk}

\author[b]{and David Wands}
\emailAdd{david.wands@port.ac.uk}

\affiliation[a]{Catholic University of Louvain - Center for Cosmology, Particle Physics and Phenomenology (CP3) 2, Chemin du Cyclotron, B--1348 Louvain-la-Neuve, Belgium}
\affiliation[b]{Institute of Cosmology and Gravitation, University of Portsmouth, Portsmouth PO1 3FX, United Kingdom}
\affiliation[c]{Institut f\"ur Theoretische Physik, Universit\"at Heidelberg, Philosophenweg 16, D--69120 Heidelberg, Germany}
\affiliation[d]{Department of Physics, Israel Institute of Technology --- Technion, Haifa 32000, Israel}
\affiliation[e]{Department of Physics and Astronomy, University of Aarhus, Ny Munkegade 120, DK--8000 Aarhus C, Denmark}

\abstract{ We show how standard Newtonian N-body simulations can be interpreted in terms of the weak-field limit of general relativity by employing the recently developed Newtonian motion gauge.
Our framework allows the inclusion of radiation perturbations and the non-linear evolution of matter. We show how to construct the weak-field metric by combining Newtonian simulations with results from Einstein--Boltzmann codes. We discuss observational effects on weak lensing and ray tracing, identifying important relativistic corrections.}

\maketitle   

\flushbottom
\section{Introduction}
\label{Introduction}

The large-scale structure that we observe in the Universe today is the result 
of the gravitational amplification of initially small fluctuations in the energy density. The dynamics is governed by the Einstein--Boltzmann equations; a set of
non-linear equations that determine the evolution of each species as well as the underlying space-time. In the standard $\Lambda$CDM cosmology,
these species are cold dark matter (CDM), baryons, massless neutrinos, photons, and
a cosmological constant ($\Lambda$).
At sufficiently early times, the linearised Einstein--Boltzmann equations approximate
the process of structure formation very accurately across a broad range of cosmological scales; second-order corrections are generally small but can be included to go beyond the linear prediction \cite{Su:2012gt,Huang:2012ub,Pettinari:2013he,Tram:2016cpy}.
This is the regime where cosmological perturbation theory \cite{Kodama:1985bj,Malik:2008im,Villa:2015ppa} is expected to provide an accurate
approximation of the underlying physical processes. 
However, at late times and small scales the non-linear collapse of matter invalidates the assumption that inhomogeneities are small with regard to the homogeneous background values and cosmological perturbation theory breaks down. 

To tackle this problem one often resorts to other tools such as cosmological N-body simulations \cite{Teyssier:2001cp,Springel:2005mi,Hahn:2015sia}, which aim to solve the non-linear gravitational collapse of matter to high accuracy. By contrast, perturbations in the radiation field, which are usually not evolved in N-body simulations (see however \cite{Brandbyge:2016raj,Adamek:2017grt}), remain sufficiently small over the whole epoch of structure formation, and may still be computed within linear perturbation theory.

In two recent papers~\cite{Fidler:2016tir,Fidler:2017ebh}, we have developed the Newtonian motion (Nm) gauge framework, which provides a relativistic space-time within which unmodified Newtonian simulations can be interpreted. Our idea is to use cosmological perturbation theory to keep track of the evolution of relativistic species and the relativistic space-time consistent with the Newtonian trajectories computed in Newtonian simulations. However, the first implementation of the Nm gauge approach relied on a dictionary that is based on linear perturbation theory, which in particular involves the assumption that the matter density perturbations are small. Although metric perturbations remain small for certain gauge choices on a vast range of scales, this implies that non-linear corrections to the metric from large matter density perturbations cannot be computed. In this paper we will remove the assumption of small matter density perturbations while keeping the smallness of metric perturbations. To achieve this in a consistent way, we propose an alternative perturbation approach based on the weak-field expansion of general relativity (GR)~\cite{Brustein:2011dy,Kopp:2013tqa,Green:2010qy,Green:2011wc,Adamek:2013wja}, and introduce the power counting scheme to perform the expansion consistently.

The rest of the paper is organised as follows. After motivating our weak-field expansion scheme in the next section, which amounts to a double expansion of all fields within the Einstein--Boltzmann system, we will introduce its leading-order equations in these expansions in section~\ref{sec:Nmtale}, where we also review the Newtonian motion gauge framework. Section~\ref{sec:Nmmetric} deals with the novel computation of the non-linear metric in the Nm framework, whereas the photon transport within the non-linear space-time is determined in section~\ref{sec:photontransport}. Relevant literature is discussed in section~\ref{sec:CZ} and we conclude 
in section~\ref{sec:conclude}. 

Notation: we use $\mu,\nu,\ldots$ to denote space-time indices and $i, j,\ldots$ for spatial indices; summation over repeated indices is assumed.

\section{Weak-field expansion scheme}

The weak-field approximation requires that gravitational potentials (corresponding to metric perturbations about the background FLRW metric in general relativity) remain small and can be thought to be of order $\epsilon \sim 10^{-5}$, a dimensionless expansion parameter. Standard cosmological perturbation theory requires in addition that every perturbation is of order $\epsilon$ or higher order, but in our weak-field limit we will allow certain matter perturbations to become large with respect to $\epsilon$. 
In the next section we will construct a consistent relativistic description in the weak-field expansion, but in this section we will first motivate our expansion scheme using the simpler Newtonian equations.

Consider the Poisson equation for the gravitational potential $\Phi$, 
\be
 \label{Poisson}
\nab^2 \Phi(\fett{x},\tau)
= -4 \pi G  \bar \rho a^2 \delta(\fett{x},\tau)
\,,
\ee
where $a$ is the cosmic scale factor, $\fett{x}$ and $\tau$ 
are conformally rescaled space-time coordinates and
$\delta(\fett{x},\tau) \equiv (\rho - \bar\rho(\tau))/\bar \rho(\tau)$ is the 
dimensionless matter density contrast.
If we assume that the potential $\Phi$ is of $\mathcal{O}(\epsilon)$ 
while the density contrast can become of $\mathcal{O}(1-100)$,  this invalidates
the underlying assumption of standard perturbation theory. A possible way out is to identify gradients with their own expansion parameter. In a situation where the density is large and the potential is small, the spatial gradient of the potential must be large. 

We therefore introduce a second expansion parameter $\kappa$ related to the spatial gradient of a perturbation. Since the gradient is dimension-full we also introduce a reference scale $k_{\rm ref}$, such that $\nab = k_{\rm ref} \mathcal{O}(\kappa)$. 
This gradient expansion is most naturally introduced in Fourier space, where the magnitude of the gradient is given by the wavenumber $k$, so that $\kappa =  k /{k_{\rm ref}}$. The form of the Poisson equation (\ref{Poisson}) 
selects the Hubble length, $H^{-1} \sim ( 4\pi G \bar \rho)^{-1/2}$, suggesting the reference scale $k_{\rm ref}=aH \equiv {\cal H}$.
Recalling that the gravitational potential $\Phi$ is of order $\epsilon$, 
we have from equation~(\ref{Poisson}) 
\be
\delta = \mathcal{O}(\kappa^2\epsilon)\,.
\ee
Next consider the Euler equation for the peculiar velocity $v_i$ of matter
\be
 \label{NEuler}
 \frac{\dd }{\dd \tau} v_i  +\Hc v_i = - \nab_i \Phi \,.
\ee
While spatial gradients may be enhanced on small scales, we will {\em not} consider cases where time derivatives become large. Thus we count 
${\cal H}^{-1} \frac{\partial}{\partial\tau} = \mathcal{O}(1)$ 
in our expansion.
Inspecting equation~\eqref{Poisson} it is evident from the Euler equation that the velocity is $\mathcal{O}(\kappa \epsilon)$.
This means that velocities may be enhanced compared to the gravitational potential, but remain small compared to the density.

Generally, the two parameters $\epsilon$ and $\kappa$ do not need to be related
and can be thought of as being separate expansion parameters that control different regimes. The former resembles the departure of perturbed quantities from their background values, whereas the latter refers to a given physical scale. 
However we can relate the two parameters in three regimes:

\begin{itemize}
\item 
$\kappa^2=\mathcal{O}(\epsilon)$.
On very large, super-Hubble scales, $k\ll {\cal H}$, we find that perturbations are nearly homogeneous and consequentially, all terms involving spatial gradients of perturbations are of order $\epsilon^2$ and may be neglected. This is the long wavelength limit describing the dynamics well beyond the horizon \cite{Lifshitz:1963ps,Tomita:1975kj,Rampf:2012pu}. In particular we see that the density contrast is of order $\epsilon^2$ in this super-Hubble limit. We will not discuss this case in detail as the physics is already well described as a limit of linear perturbation theory.
\item
$\kappa^2=\mathcal{O}(1)$.
The second range of scales is such that $\kappa = \mathcal{O}(1)$. Here spatial gradients are neither enhanced nor suppressed and we recover standard perturbation theory. For our analysis we will employ the resulting linear equations from very large scales up to the non-linear scales where $\kappa$ becomes large and standard perturbation theory eventually breaks down. 
\item
$\kappa^2=\mathcal{O}(\epsilon^{-1})$.
Finally, on small scales $\kappa$ is large and we must solve the non-linear dynamics. However, while the density grows, we assume that the gravitational potentials remain small. We may exploit that dynamically $\delta = \mathcal{O}(\kappa^2\epsilon) = \mathcal{O}(1)$ holds to link the two expansion parameters and conclude that $\kappa = \mathcal{O}(\epsilon^{- 1/2})$. This remains valid on intermediate linear and non-linear scales, but not at the very small scales where strong gravity may become relevant (e.g.\ for black-hole formation).

\end{itemize}

Starting from the fully non-linear system of Einstein--Boltzmann equations, we keep only the leading-order terms in $\epsilon$ (while assuming $\kappa = \mathcal{O}(1)$) plus the leading terms when counting $\kappa = \mathcal{O}(\epsilon^{-1/2})$. 
In the scalar sector, this expansion scheme amounts to keeping only the linear (or leading-order) terms in the metric/gravitational potentials,  
but the source of these perturbations, i.e., the matter fluid variables, is incorporated in full non-linearity.\footnote{In this spirit and for simplicity,  we will often abbreviate ``non-linearly sourced metric potentials'' with ``non-linear metric potentials''.}
This way, we obtain a unified scheme that encompasses the largest scales where linear perturbation theory is an accurate description, and the leading contributions from the non-linear regime.

Our double expansion scheme is in agreement with the one as employed in \cite{Brustein:2011dy,Kopp:2013tqa}, whereas it differs in the treatment of subleading terms (according to our counting) from the one of \cite{Green:2010qy,Green:2011wc,Adamek:2013wja,Eingorn:2015hza,Eingorn:2016kdt,Brilenkov:2017gro,Goldberg:2016lcq,Goldberg:2017gsm}. All expansions agree for the leading terms.
To the leading order and for non-relativistic matter, our expansion scheme further agrees with the one as employed in the post-Friedmann approach \cite{Milillo:2015cva,Rampf:2016wom}.

\section{Newtonian motion gauges}\label{sec:Nmtale}

\subsection{Notation and conventions} \label{sec:notdef}

We define metric perturbations about a homogeneous and isotropic Friedmann--Lema\^itre--Robertson--Walker background as follows without specifying the gauge 
\begin{subequations}
\label{metric-potentials}
\begin{align}
  g_{00} &= -a^2 \left( 1 + 2 A \right) \,, \\
  g_{0i} &=  -a^2 \left(B_i + \hat\nab_i B\right) \,,\\
  g_{ij} &= a^2 \left[ \delta_{ij} \left( 1 + 2 \HL \right) + 2 \left(  \hat\nab_i \hat\nab_j + \frac {\delta_{ij}}  {3} \right) \HT - \hat\nab_i H_{{\rm T}j} - \hat\nab_j H_{{\rm T}i} - 2 H_{{\rm T}ij} \right] \,.
\end{align}
\end{subequations}
We have defined a normalised gradient operator $\hat\nab_i \equiv -(-\Delta)^{-1/2} \nab_i$ where $\Delta = \nabla^2$ is the Laplacian; furthermore we have the perturbations 
$A$ and $B$ of the scalar type, $B_i$ and $H_{{\rm T}i}$ of the vector type satisfying $\hat\nab^i B_i = 0 = \hat\nab^i  H_{{\rm T}i}$, and the trace-free tensor $H_{{\rm T}ij}$ that is constrained with $H_{{\rm T}i}^i = 0 = \hat\nab^i H_{{\rm T}{ij}}$. The operator $\hat\nab_i$ is also known as the Riesz transformation, and is replaced by $-\ii \hat{k}_i$ in Fourier space, where $\hat k_i \equiv k_i/|\fett{k}|$.

One immediate advantage of our notation is that the scalar potentials exactly match the usual Fourier space definitions for example as employed in \cite{Blas:2011rf}. The ordinary gradient operator is $\nab \sim \mathcal{O}(\kappa)$, while the normalised gradient operator is of order one. We further define the operator $\kop = (-\Delta)^{1/2} $, which in Fourier space becomes the magnitude $k = |\fett{k}|$
and is of order $\kappa$. These operators greatly simplify the order counting in the subsequent sections. Note in particular that the power counting of the metric perturbations within the weak-field limit is simply that all metric perturbations ($A$, $\HL$, $B$, $B_i$, $\HT$, $H_{{\rm T}i}$ and $H_{{\rm T}ij}$) are of \mbox{order $\epsilon$.}

The matter and radiation content is characterised by the density, pressure, velocity and anisotropic stress, which are defined in the fluid rest frame.
We decompose them into a scalar density $\rho=\bar\rho(1+\delta)$, pressure $p$, scalar velocity $v$, vector velocity $v^i$ and the scalar stress $\Sigma$, vector stress $\Sigma^i$ and tensor stress $\Sigma^{ij}$.
The energy-momentum tensor is given by
\begin{subequations}
\label{enegry-momentum}
\begin{align}
  T^0_{\phantom{0}0} &= - \rho \,, \\
 T^0_{\phantom{0}i} &= (\rho+ p) (v_i + \hat\nab_i v -B_i - \hat\nab_i B) \,, \\ 
 T^i_{\phantom{i}j} &= p \delta^i_j  + \left(  \hat\nab^i \hat\nab_j + \frac {\delta^i_{j}} {3} \right) \Sigma - \frac 1 2 \left(\hat\nab^i \Sigma_j + \hat\nab_j \Sigma^i \right) - \Sigma^i_{\phantom{i}j} 
 + (\rho+p)(v^i + \hat\nab^i v)(v_j + \hat\nab_j v) \,,
\end{align}
\end{subequations}
and is composed of the sum over the energy-momentum tensors of all individual species 
(matter, photons and massless neutrinos), while the total energy momentum tensor is the sum over the energy-momentum tensors of all species.
The corrections proportional to the velocity squared to the anisotropic stress result from its definition in the fluid rest frame.

The weak-field expansion orders of the matter perturbations are derived from the dynamical equations. We have already argued, employing the Poisson equation, that the density may be large and $\delta= \mathcal{O}(\kappa^2 \epsilon)$. Velocities are also enhanced but remain small compared to the density and we find $v^i = \mathcal{O}(\kappa \epsilon)$ and $v = \mathcal{O}(\kappa\epsilon)$. The anisotropic stress is generally small and for simplicity we assume that it is of order $\epsilon$, i.e. $\Sigma \sim \Sigma^{i} \sim \Sigma^{ij} = \mathcal{O}(\epsilon)$. A priori we cannot know if this assignment of orders is self-consistent, but this can be checked explicitly once we write down the complete set of equations.

To determine the weak-field equations, we identify the leading small-scale terms (counting $\kappa = \epsilon^{- 1/2}$) in every equation and the leading Hubble-scale terms (counting $\kappa = 1$) and neglect all remaining 
sub-leading terms (e.g., non-linear combinations of the gravitational potentials). This provides ``non-linear'' equations that capture the physics on both large and small scales accurately.

\subsection{The temporal gauge condition}

The temporal gauge condition does not affect the velocities of the particles to leading order in $\epsilon$ and as such it is not directly related to the concept of Nm gauges. However, some temporal gauge choices may be inconsistent with our weak-field expansion for the metric.
For example, in a comoving gauge we require $v=B$ which implies that $B= \mathcal{O}(\kappa\epsilon)$, violating our assumption that $B= \mathcal{O}(\epsilon)$. 

We therefore adopt a temporal gauge condition which coincides with the Poisson- or longitudinal-gauge time coordinate 
\be 
 \label{temporalgauge}
\kop B =  \dot{H}_{\rm T}\,,
\ee
where the operator $\kop$ is defined in section~\ref{sec:notdef}.
The metric potential $\HT$ will be uniquely defined by our spatial gauge and if it is of order $\epsilon$, then it follows that the shift vector becomes even smaller ($B = \mathcal{O}(\kappa^{-1}\epsilon)$). However, this does not invalidate our more general expansion of $\delta g_{\mu\nu} = \mathcal{O}(\epsilon)$.
Note that this temporal gauge choice differs from the one used in our recent papers~\cite{Fidler:2015npa,Fidler:2016tir,Fidler:2017ebh}.

\subsection{The spatial gauge condition}

The spatial gauge is fixed by the Nm gauge condition, which demands that the relativistic Euler equation for matter takes the Newtonian form. Before formulating 
the spatial gauge condition we first need to investigate the relativistic equations in the weak-field limit employing the temporal gauge condition~\eqref{temporalgauge}.

We derive the weak-field dynamical equations 
using the software package {\sc XPand} \cite{Pitrou:2013hga} and employ these to derive the spatial gauge condition that guarantees Newtonian motion. For the relativistic Poisson equation in the weak-field limit we obtain
\be
 \label{rel:Poisson}
4\pi G a^2 \left(\delta\rho + 3 \Hc(\rho+p)\kop^{-1}\left( v - \kop^{-1} \dot{H}_{\rm T} \right)\right) = \kop^2 \left(\HL + \frac 1 3 \HT \right) \,,
\ee
where we define the density and pressure perturbations using $\rho = \bar{\rho} + \delta\rho$ and $p = \bar{p} +\delta p$.

The Einstein equations are complemented by the energy and momentum conservation equations, valid for the total perturbations and for each uncoupled component of the Universe separately. Energy conservation provides the relativistic continuity equation 
\be
\left(\partial_\tau + 3 \Hc\right)\delta\rho + 3 \Hc \delta p = -3 (\rho+p)\dot{H}_{\rm L} + (v^i+ \hat\nab^i v)\hat\nab_i \kop \rho - (\rho+p) \kop v \,.
\ee
The volume deformation $\HL$ that appears in the relativistic continuity equation above leads us to define the coordinate or counting density 
\be
 \rho_{\rm count} = \rho + 3(\rho + p)\HL \,.
\ee
The counting density is by definition based only on the particle positions, while the relativistic density describes the physical energy content. The connection between both can be derived as a geometrical effect. In general relativity the physical volume in a coordinate cell depends on the trace of the (conformally rescaled) spatial metric $(1+ 2 \HL)$,
introducing a mismatch of $3\HL \rho$ between both densities. The second contribution, $3\HL p$, is relevant for relativistic species; since they exert a pressure, the energy density is further changed by a local volume perturbation. In addition to having the particles concentrated in a smaller volume, the gas is compressed, further increasing the energy density.   

At the level of the density contrast we may write 
\be 
 \delta_{\rm count} = \delta + \frac{3 (\bar\rho + \bar p)}{\bar \rho} \HL  \,. \label{eq:countdens}
\ee
Here we have neglected terms of the type $\HL \delta$; while the density contrast may become large, it turns out that this correction is sub-leading on all relevant scales. On the large scales it is of order $\epsilon^2$ and may be neglected compared to the leading linear terms. On the small scales, the density contrast is of order one, while terms like $\HL \delta$ are only of order $\epsilon$. In both cases the additional term is suppressed by one order of $\epsilon$ compared to the leading terms and thus may be neglected. 

For massive, non-relativistic species we may thus employ $\delta_{\rm count} = \delta + 3 \HL$ and for relativistic species $\delta_{\rm count} = \delta + 4 \HL$. 

In the case of cold dark matter ($p^{\rm cdm} = 0$), to leading order in our weak-field expansion, we obtain 
\be
\dot{\delta}_{\rm count}^{\rm cdm}  =   (v_{\rm cdm}^i+ \hat\nab^i v_{\rm cdm})\hat\nab_i \kop \delta_{\rm count}^{\rm cdm} - (1+\delta_{\rm count}^{\rm cdm}) \kop v_{\rm cdm}\,,
\ee 
which has the same form as the non-linear but Newtonian continuity equation.

The momentum constraint provides the relativistic Euler equation 
\begin{align} \nonumber
\left(\partial_\tau +\Hc\right)(v_i &+ \hat\nab_i v -  B_i - \kop^{-1} \hat\nab_i  \dot{H}_{\rm T}) - (v^j + \hat\nab^j v) \hat\nab_j \kop ( v_i + \hat\nab_i v) - \hat\nab_i \kop A  \\ 
&=  -\frac{1}{\rho+p}\left(\dot{p}(v_i + \hat\nab_i v  - B_i - \kop^{-1} \hat\nab_i \dot{H}_{\rm T}) - \hat\nab_i \kop \delta p + \frac 2 3 \hat\nab_i \kop \Sigma - \frac 1 2 \kop \Sigma_i\right) \,,
\end{align}
which for cold dark matter simplifies to 
\begin{align} \nonumber
\left(\partial_\tau +\Hc\right)(v^{\rm cdm}_i + \hat\nab_i v^{\rm cdm} & - B_i - \kop^{-1}\hat\nab_i \dot{H}_{\rm T} ) - ( v_{\rm cdm}^j + \hat\nab^j v^{\rm cdm} ) \hat\nab_j \kop(v^{\rm cdm}_i + \hat\nab_i v^{\rm cdm}) -\hat\nab_i \kop A  \\
&=  -\frac{1}{\rho}\left( \frac 2 3 \hat\nab_i \kop \Sigma^{\rm cdm} - \frac 1 2 \kop \Sigma^{\rm cdm}_i\right) \,.
 \label{Eulercdm}
\end{align}
The Newtonian motion (Nm) gauge condition is now derived from the requirement that this relativistic equation for cold dark matter (\ref{Eulercdm}) has the same form as the Newtonian Euler equation. The non-Newtonian contributions are the appearance of the metric potentials $\HT$, $B_i$ and $A$, while the Newtonian Euler equation is instead sourced only by the Newtonian potential $\Phi^{\rm N}$.
Thus for scalar perturbations we find the Nm gauge condition 
\be
 \label{def:Nmgauge}
A + \left(\partial_\tau +\Hc\right) \kop^{-2} \dot{H}_{\rm T}   = -\Phi^{\rm N} \,,
\ee 
where the Newtonian potential obeys the Newtonian Poisson equation \be
4\pi G a^2 \delta\rho_{\rm count}^{\rm cdm} = \kop^2 \Phi^{\rm N}\,,
\ee
which is sourced by the counting density of the massive species. 
Equation~\eqref{def:Nmgauge} coincides with the spatial Newtonian motion gauge condition defined in~\cite{Fidler:2016tir,Fidler:2017ebh}, but now 
the metric potentials incorporate non-linear sources.

The Newtonian theory has no gravitational vector potential, and vector modes of the velocity are only generated dynamically. 
The Newtonian motion gauge condition in the vector sector is thus eliminating the vector potential $B_i$ by setting $B_i = 0$. 

Note that in both the scalar and the vector sector, these gauge conditions do not completely fix the gauge, but leave open boundary conditions. These are connected to the initial velocity and density in the simulation.
For more details see the discussion in section 4 of~\cite{Fidler:2016tir}.

Inserting the Newtonian motion gauge condition (\ref{def:Nmgauge}) and $B_i = 0$ into the cold dark matter Euler equation (\ref{Eulercdm}), we recover the Newtonian Euler equation
\begin{align}
\left(\partial_\tau +\Hc\right)(v^{\rm cdm}_i + \hat\nab_i v^{\rm cdm}) &- (  v_{\rm cdm}^j + \hat\nab^j v^{\rm cdm}) \hat\nab_j \kop( v^{\rm cdm}_i + \hat\nab_i v^{\rm cdm}) +\hat\nab_i \kop \Phi^{\rm N}  \nonumber \\
&=  -\frac{1}{\rho}\left( \frac 2 3 \hat\nab_i \kop \Sigma^{\rm cdm} - \frac 1 2 \kop \Sigma^{\rm cdm}_i\right) \,.
\end{align}
The appearance of the anisotropic stress in equation~\eqref{Eulercdm} is in fact a Newtonian term and describes the impact of a non-trivial phase space on the dark matter evolution; 
in Newtonian theory the anisotropic stress is defined as the second kinetic moment  of the particle distribution function (see e.g.\ \cite{Bernardeau:2001qr}).

It follows that the subset of perturbations $\rho^{\rm cdm}_{\rm count}$, $v^{\rm cdm}$, $v_i^{\rm cdm}$ and $\Phi^{\rm N}$, as well as the anisotropic stress extracted from the particle representation $\Sigma^{\rm cdm}$, $\Sigma^{\rm cdm}_i$, $\Sigma^{\rm cdm}_{ij}$ follow Newtonian and non-linear equations of motion and may be accurately computed in a Newtonian N-body simulation. We therefore identify them with their counterparts in such a simulation, labeled with a superscript 'N'
\be \label{ident}
 \rho_{\rm cdm}^{\rm count} = \rho^{\rm N} \,, \;\; v^{\rm cdm} = v^{\rm N}\,, \;\; v_i^{\rm cdm}= v^{\rm N}_i\,,  \;\; \Sigma^{\rm cdm} = \Sigma^{\rm N} \,,  \;\; \Sigma_i^{\rm cdm} = \Sigma^{\rm N}_i \,,  \;\; \Sigma_{ij}^{\rm cdm} = \Sigma^\text{N}_{ij}\,.
\ee

\section{Computing the Newtonian motion gauge metric}\label{sec:Nmmetric}

In the preceding section we have shown how we can define relativistic weak-field equations that coincide with the standard non-linear Newtonian equations of motion used to evolve matter. In this section, we will discuss how to construct the corresponding metric in GR and how relativistic species can be evolved.

Note that our metric potentials go beyond first-order perturbations calculated in standard perturbation theory, which we denote with a superscript $(1)$. 
The linear potential, $A^{(1)}$, is unaffected by the non-linear small-scale growth of matter perturbations, while our weak-field potential, $A$, remains small but is sourced by the non-linear matter density. Thus we cannot use an ordinary linear Einstein--Boltzmann code to compute $A$ to the desired weak-field precision. 

However, we know that on the large scales and at early times the linear approximation is accurate and this allows us to replace some weak-field perturbations by their linear counterparts computed for example in \CLASS~\cite{Blas:2011rf}. 

In particular we can use linear theory to calculate the radiation perturbations. These only affect the metric in the early Universe and remain well-described by the linear approximation on all scales. Thus we identify $\delta_\gamma = \delta_\gamma^{(1)}$, $v_\gamma = v_\gamma^{(1)}$ and $\Sigma_\gamma = \Sigma_\gamma^{(1)}$. Together with the non-linear dark matter density, evolved in N-body simulations, this provides an accurate description in a standard model cosmology for all the fluids in the weak-field approximation.

For the computation of the metric perturbations from the Einstein equations we will first define (where quantities in an arbitrary gauge are denoted by tildes)
\be 
\Phi \equiv \widetilde \HL + \frac 1 3 \widetilde \HT + \Hc \kop^{-1} \left( \widetilde B - \kop^{-1} \dot{\widetilde{{H}_{\rm T}}}  \right) =\HL + \frac 1 3 \HT \,, \label{eq:PhiDef}
\ee
which at linear order reduces to the Bardeen potential and where we have applied the temporal gauge choice (\ref{temporalgauge}) in the second equality. Inserting equation~\eqref{eq:PhiDef} in the trace-free part of the weak-field Einstein equations we get
\begin{align}
A + \HL  + \frac 1 3 \HT  = A + \Phi  &= -8 \pi G a^2 \kop^{-2}(\Sigma + \Sigma^\text{GR})\,. \label{sigmaeq}
 \end{align}
In this equation we have introduced the shorthand notation $\Sigma^\text{GR}$ for terms that are quadratic in the metric perturbations and thus of order $\epsilon^2$. These are sub-leading in our expansion and are only needed to compute suppressed quantities such as the difference of the two scalar potentials, where these are now of leading order in the case that the anisotropic stress $\Sigma$ is suppressed. See~\cite{Adamek:2017grt} for an example of $\Sigma^\text{GR}$ in the Poisson gauge. From the energy constraint equation~(\ref{rel:Poisson}) we find
\begin{align}
4\pi G a^2 (\delta\rho  + 3 \Hc  (\rho+p) \kop^{-1}(v - \kop^{-1}\dot{H}_{\rm T})) &= \kop^2 \Phi\,.
\end{align}
The remaining metric potentials are constrained by the Newtonian motion gauge conditions (\ref{temporalgauge}) and (\ref{def:Nmgauge}) 
\begin{align}
 \dot{H}_{\rm T} &= \kop B\,, \\ \label{eq:Nmgaugedef}
-\left(\partial_\tau +\Hc\right) \kop^{-2} \dot{H}_{\rm T} &= A+\Phi^{\rm N}\,.
\end{align}
On large scales ($\kappa \sim 1$) the dominant terms in equation~\eqref{eq:Nmgaugedef} are linear and all contributions are equally important.
However, on small scales ($\kappa\gg1$), the right hand side appears to be of order $\epsilon$, while the left hand side is only of order $\kappa^{-2}\epsilon$. 
If there is no cancellation between the potentials on the right, then our spatial gauge condition would violate the assumed weak-field orders. The reason is that the impact of $\HT$ on the dark matter motion is suppressed on the small scales by a factor of $\kop^{-2}$ and any mismatch in the dark matter motion needs to be compensated with an increasingly large value of $\HT$.

Fortunately, to leading order on small scales, $A$ is almost identical to $-\Phi$ since the anisotropic stress in equation~\eqref{sigmaeq} is equally suppressed by $\kop^{-2}$. In addition $\Phi$ and $\Phi^{\rm N}$, obey an almost identical Poisson equation and we find that $A +\Phi^{\rm N}$ is in fact of order $\kappa^{-2}\epsilon$ as expected. 

The spatial gauge condition is thus consistent with the assumption that $\HT =\mathcal{O}(\epsilon)$. 
Explicitly evaluating the gauge condition~(\ref{eq:Nmgaugedef}) we find
\be
(\partial_\tau +\Hc)\dot{H}_{\rm T} = 4\pi G a^2 (\delta\rho_{\gamma} +3 \Hc (\rho_{\gamma} + p_{\gamma})\kop^{-1} (v - \kop^{-1}\dot{H}_{\rm T}) - \rho_{\rm cdm}(3\zeta -  \HT)) + 8 \pi G a^2 \Sigma\,,
 \label{gaugedefinition}
\ee
where we have used the label $\gamma$ to indicate the combined density or pressure of all relativistic species, and made use of the comoving curvature perturbation $\zeta$, defined through
\be
 \zeta \equiv \widetilde \HL + \frac 1 3  \widetilde \HT - \Hc \kop^{-1}(\widetilde v - \widetilde B) \,.
\ee
In the limit when relativistic species and anisotropic stress are negligible,
we recover the result from linear theory that the choice $\HT = 3\zeta$ delivers a stable solution to the metric.
On the small scales the anisotropic stress of the massive species may contribute to the evolution of $\HT$. However, the impact of $\HT$ on the dark matter motion on these scales is suppressed and decouples from the leading order as long as $\HT$ remains of order $\epsilon$. Therefore we may substitute $\Sigma \rightarrow \Sigma_{\gamma}$ in our gauge definition~(\ref{gaugedefinition}), which includes the important large-scale contribution from radiation, but not the small-scale anisotropic stress of matter that has a negligible impact on the dark matter motion.
This choice of gauge provides a particularly simple Newtonian motion gauge that describes the dark matter motion accurately in the weak field limit.
The source for $\HT$ is now composed of only radiation perturbations and $\zeta$, which both may be computed accurately from linear theory. It follows that $\HT = \HT^{(1)}$ and the non-linear small-scale physics does not affect it. 

We therefore conclude that while most metric potentials receive corrections from the non-linear dynamics, $\HT$ and $B$ are accurately described by linear physics (and $B^i$ too, but it is zero because of our gauge conditions). We may employ \CLASS{} to compute these metric potentials. The remaining potential $A$ on the other hand is affected by the non-linear density and related to the full simulation potential $\Phi^{\rm N}$ via the spatial gauge condition $\Phi^{\rm N}+A=-\left(\partial_\tau +\Hc\right) \kop^{-2}\dot{H}_{\rm T}$. The non-linear spatial potential $\HL$ is connected to $A$ via the metric constraint~\eqref{sigmaeq}. 

\subsection{GR dictionary}\label{sec:dic}

In this section we summarise the connection between a Newtonian simulation and
the weak-field limit of GR. We have shown in the previous sections that a Newtonian simulation is computing $\delta^{\rm N}$, $v^{\rm N}$ and $\Phi^{\rm N}$ in the non-linear weak-field limit of GR. We may embed the simulation into a Newtonian motion gauge space-time with the metric potentials $\HT$ and $B$ computed in a linear Boltzmann code. The remaining relativistic perturbations may be obtained from
\begin{align}\label{eq:NMA}
A &= -\Phi^{\rm N} - \left(\partial_\tau + \Hc\right) \kop^{-2}\dot{H}_{\rm T} \,,\\\label{eq:NMHL}
\HL &=   \Phi^{\rm N}  - \frac 1 3  \HT - \gamma \,,\\\label{eq:NMv}
v &= v^{\rm N}\,, \\\label{eq:NMd}
\delta &= \delta^{\rm N} - 3\HL =  \delta^{\rm N} - 3 \Phi^{\rm N}   + \HT + 3 \gamma \,,
\end{align}
where we have introduced the quantity $\gamma$ defined by
\begin{equation}
\gamma \equiv - \left(\partial_\tau + \Hc\right)\kop^{-2} \dot{H}_{\rm T} + 8 \pi G a^2 \kop^{-2}\Sigma,
\end{equation}
which we have employed in our previous papers~\cite{Fidler:2016tir,Fidler:2017ebh} and describes the impact of relativistic species on the dark matter motion. 
Note that the Nm gauge condition~\eqref{def:Nmgauge} can be expressed as 
\begin{equation} \label{altdef:Nmgauge}
\Phi^\text{N} - \Phi = \gamma\,,
\end{equation}
by using the trace-free part of the Einstein equation~\eqref{sigmaeq}.

These equations are sufficient to construct all relativistic perturbations from the combined output of the N-body simulation ($\Phi^{\rm N}$, $\rho^N$, $v^{\rm N}$ and $\Sigma^{\rm N}$) and a linear Boltzmann code ($\HT$, $\gamma$ and the relativistic species). Further note that at the small scales the density, velocity and lapse perturbation decouple from the potential $\HT$ to leading order and are uniquely determined from the simulation alone.   

In the late-time limit of a radiation free Universe, the choice of $\HT = 3\zeta$ provides a particularly simple weak-field Newtonian motion gauge. The temporal gauge condition states that $B$ vanishes in this case. The remaining potential $A = -\Phi^{\rm N}$ is the non-linear simulation potential, while $\HL =  \Phi^{\rm N} - \frac{1}{3} \HT$.  

\begin{figure}[tb]
   \centering
     \includegraphics[width=\textwidth]{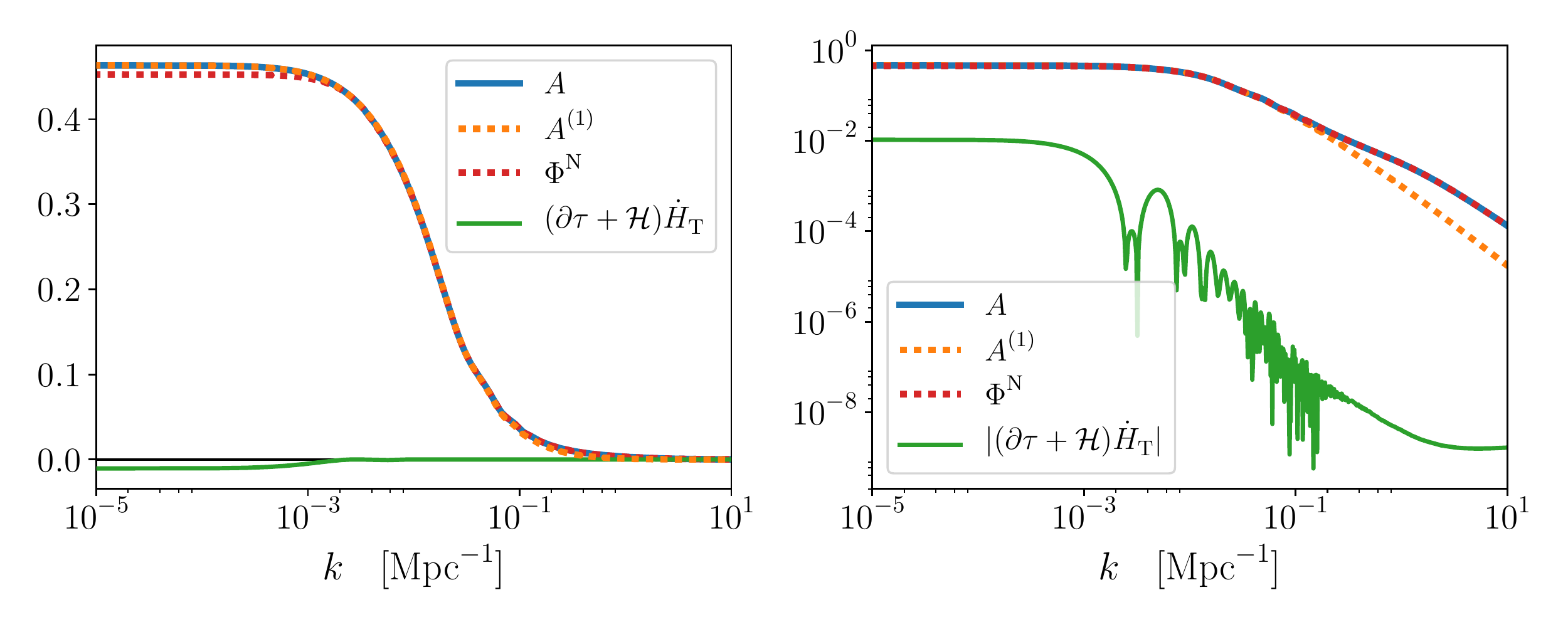}
   \caption{We show an example of the construction of the non-linear lapse from the simulation potential $\Phi^\text{N}$, which is dominant on the small scales, and the evolution of $\HT$ due to radiation, which is accurately described in linear theory and adds a correction on the larger scales. For comparison we also plot the linear lapse $A^{(1)}$ that would be obtained in a linear Nm gauge. For illustration purposes only, the non-linear potential $\Phi^\text{N}$ has been estimated using \textsc{Halofit}~\cite{Takahashi:2012em} as $\Phi^\text{N} \simeq \Phi^\text{N, linear} \times \sqrt{\frac{P(k)^\text{Halofit}}{P(k)^\text{linear}}}$. All transfer functions are normalised to $\zeta=-1$ at super-horizon scales which is the default in \CLASS{}.}
   \label{fig:Apot}
 \end{figure}

We construct the non-linear metric from the combined output of \CLASS{} and a Newtonian simulation. In Figure~\ref{fig:Apot} we show the lapse $A$, as computed from $\Phi^\text{N}$ and $\HT$ and compare it against the linear version of $A^{(1)}$ that we would obtain using only a linear Nm gauge instead of our new weak-field approach. Apart from the changed temporal gauge these represent the results in our previous works \cite{Fidler:2016tir,Fidler:2017ebh}. We find that the lapse perturbation on the small scales is dominated by $\Phi^\text{N}$ and significantly deviates from the linear $A^{(1)}$. On the large scales corrections from $\HT$ become relevant and it differs from the Newtonian potential, while the linear computation from \CLASS{} is very accurate. 

\paragraph{Relations in Poisson gauge.}
It is often convenient to have an analogous dictionary for the Poisson gauge. Since the present choice of Nm gauge and the Poisson gauge share the same temporal gauge, we immediately have
\begin{align} \label{eq:Pphi}
\Psi \equiv A^{\rm P} &= A = -\Phi^{\rm N}  - \left(\partial_\tau + \Hc\right) \kop^{-2}\dot{H}_{\rm T}\,, \\ 
\delta^{\rm P} &= \,\delta \,= \delta^{\rm N} - 3 \Phi^{\rm N}  + \HT + 3\gamma \,. \label{eq:densdicfull}
\end{align}
The other scalar potential in the Poisson gauge is given by equation~\eqref{altdef:Nmgauge} as
\be \label{eq:Ppsi}
\Phi \equiv \HL^{\rm P} = \Phi^\text{N} - \gamma\,.
\ee 
Finally, the velocity in Poisson gauge is not identical to the Nm gauge velocity and we obtain it from the spatial gauge transformation connecting both gauges
\be\label{eq:Pv}
v^{\rm P} = v^{\rm N} - \kop^{-1}\dot{H}_{\rm T}\,.
\ee
\begin{figure}[tb]
   \centering
     \includegraphics[width=\textwidth]{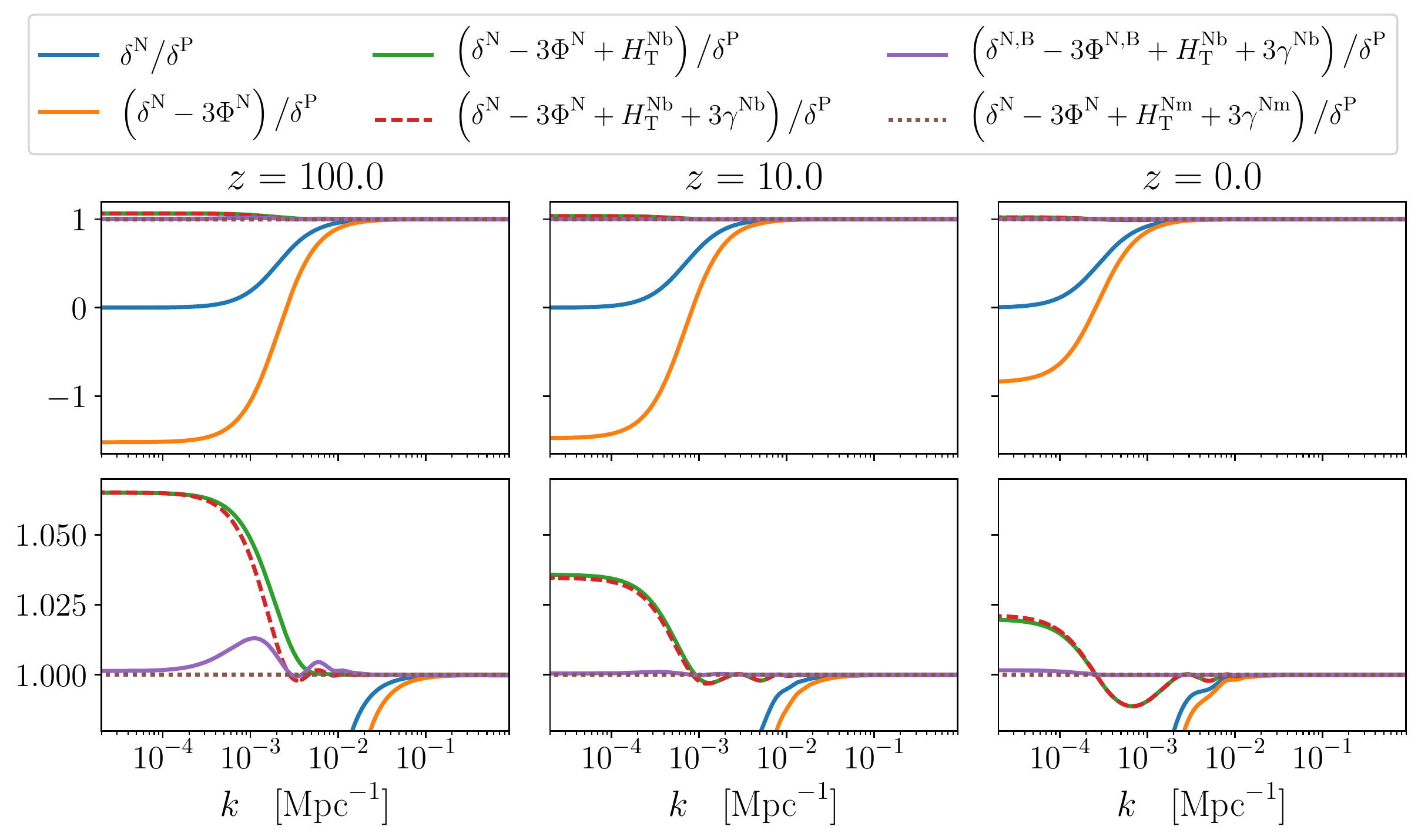}
     \caption{We show different approximations to the Poisson gauge density; see equation~(\ref{eq:densdicfull}) and surrounding text for further details.
The abbreviation ``N,B'' refers to a quantity of a Newtonian simulation using backscaled initial conditions (as discussed in section~\ref{sec:BS}), whereas ``N'' refers to the Newtonian perturbations when not employing backscaling. ``Nb'' refers to quantities within the N-body gauge \cite{Fidler:2015npa}.
}
   \label{fig:dirupt}
\end{figure}

In figure~\ref{fig:dirupt} we show the relative importance of the various contributions to the Poisson gauge density~\eqref{eq:densdicfull} as an example of using our dictionary. On large scales, $\delta^\text{N}$ is very different from $\delta^{\rm P}$ due to the volume deformation term $-3\HL$ in equation~\eqref{eq:countdens}. A naive (wrong) gauge transformation of equation~\eqref{eq:densdicfull} (i.e. ignoring the spatial gauge generator and thus the non-trivial space-time of the simulation) leads to the approximation $\delta^\text{P} \approx \delta^{\rm N}-3\Phi^\text{N}$, which is even worse than neglecting relativistic effects altogether and assuming $\delta^\text{P} \approx \delta^{\rm N}$. The following approximation $\delta^\text{P} \approx \delta^{\rm N} - 3 \Phi^{\rm N}  + \HT$, includes the important geometrical impact of $\HT$ and reduces the error to the \%-level. This line is equivalent to an analysis of the simulation in the N-body gauge, which would be accurate in the absence of radiation. The next two lines add the correction from radiation due to $\gamma$, but still evaluate the metric in the N-body gauge. This turns out to be a very good approximation in the case of a backscaling simulation (discussed in section~\ref{sec:BS} and providing $\delta^{\rm N,B}$, $v^{\rm N,B}$ and $\Phi^{\rm N,B}$) as the metric remains close to the N-body gauge for most of the evolution. The final line is employing the full Nm gauge metric and finds an excellent agreement with the Poisson gauge density.

In summary, we see that on the large scales, the contributions from $\Phi^\text{N}$ and $\HT$ are of the same order of magnitude and partially cancel. Both are required for a consistent interpretation of a Newtonian simulation, with the remaining terms taking into account the impact of radiation. 

\subsection{Backscaling}\label{sec:BS}
Initial conditions for Newtonian N-body simulations which do not include radiation perturbations are usually set by rescaling the present-day matter power spectrum obtained from linear Einstein--Boltzmann codes back to the desired initial redshift; a procedure known as backscaling. 

For a $\Lambda$CDM Universe, backscaling has been shown to work very well in the linear case~\cite{Fidler:2017ebh}, and we now extend the backscaling method to the weak-field limit. In the Nm gauge framework, backscaling amounts to fixing the spatial gauge freedom at the final time by enforcing $ \HT = 3\zeta$ at that time. Then, by definition, the present day simulation density must match the comoving density. Since we are employing a Newtonian motion gauge, we may further employ Newtonian equations of motion for the simulation density. Thus the initial conditions may be found by scaling the present day dark matter distribution back to the initial time using the Newtonian growing mode solution. This provides the well known backscaling initial conditions utilised in most N-body simulations.

Using backscaling initial conditions, the Nm gauge metric remains trivial throughout most of the cosmic evolution and only deviates from $\HT = 3\zeta$ at the early times when relativistic species are relevant. As a consequence we may exploit a particularly simple relation between the simulation and general relativity in the late Universe based only on the simulation output and the comoving curvature perturbation $\zeta$: 
\be 
\delta = \delta^{\rm N} - 3 \left(\Phi^{\rm N}  + \zeta \right) \,, \;\;
v = v^{\rm N}\,, \;\;
A = -\Phi^{\rm N}  \,, \;\;
B= 0 \, , \;\;
\HL =  \Phi^{\rm N}  + \zeta  \,, \;\;
\HT = 3\zeta \,.
\label{backscaling}
\ee
Note that assuming linear perturbation theory, $\zeta$ may also be extracted from the simulation output directly by employing the relation
\be \label{eq:zeta}
\zeta = - \Hc \kop^{-1} v^{(1)} - A^{(1)} \approx -\Hc \kop^{-1} v^{\rm N} + \Phi^{\rm N} \,.
\ee
However, the simulation perturbations do not remain linear and thus this relation might produce incorrect results. To be completely consistent $\zeta$ should be extracted from a linear Boltzmann code. 

\subsection{Vector and tensor perturbations}\label{sec:vec}

So far we have only discussed the dominant scalar sector.
The vector gauge condition is simply $B^i= 0$, while there is no tensor gauge condition. 
The remaining metric degrees of freedom are constrained by the Einstein equations
\begin{align} \label{eq:vec}
\left(\partial_\tau + 2\Hc\right)\dot{H}_{{\rm T}i} &= 8 \pi G a^2 (\Sigma_i + \Sigma^\text{GR}_i) \\
\left(\partial_\tau + 2\Hc\right)\dot{H}_{{\rm T}ij} + \kop^2 H_{{\rm T}ij} &= 8 \pi G a^2 (\Sigma_{ij} + \Sigma^\text{GR}_{ij})  \,. \label{eq:ten}
\end{align}
These are the usual decaying mode for vectors and a damped oscillation for tensors (including gravitational waves). Both are sourced by the anisotropic stress and higher-order corrections from GR (within our expansion), which we list here only since they may be the dominant sources is the absence of an anisotropic stress.

In order to solve these equations, we require two additional boundary conditions for the vector $\HT^i$ and four for the tensor $\HT^{ij}$. The vector degrees of freedom are related to the initial Newtonian vector velocities $v^{{\rm N},i}$, typically assumed to be zero, just as the scalar degrees of freedom are related to the initial Newtonian density and scalar velocity. The tensor degrees of freedom are not related to our gauge choice and are simply the initial dynamical degrees of freedom for the gravitational waves. 

Since the tensor and vector metric perturbations do not affect the leading-order dark matter equations of motion in a Nm gauge, we may integrate them separately from the N-body simulation, employing its output. Their feedback on the dark matter simulation is a higher-order effect and may be neglected.  

The tensor equation on the small scales does not describe dynamical waves, but a displaced oscillator at rest. The dominant term on the right hand side is given by $\kop^2 H_{{\rm T}ij}$ while the time derivatives may be neglected. Since the source on the left-hand side is small, this leads to an additional suppression and we find that $H_{{\rm T}ij} = \mathcal{O}(\kappa^{-2}\epsilon^2)$ and thus much smaller than our initial weak-field assumption. $\HT^i$ on the other hand is decaying up to a late time residual of order $\mathcal{O}(\kappa^{-1}\epsilon^2)$.

We conclude that the vector and tensor modes of the metric are strongly suppressed on the small scales, while they are given by the standard linear results on the larger scales. This is in agreement with the results found in~\cite{Adamek:2015eda}.

\section{Photon transport}\label{sec:photontransport}

An important addition to relativistic particle simulations is the computation of the photon transport to construct gauge invariant observables. Since the Nm gauge framework provides the non-linear metric it can be used to perform this task consistently within weak-field approximations. 

We employ the well established results in the literature based on linear perturbation theory \cite{Yoo:2009au,Bonvin:2011bg,Challinor:2011bk,Jeong:2014ufa}. In the weak-field limit, the metric potentials are non-linear, but we may still exploit that they are small and terms proportional to the metric squared may be neglected as long as there are not enough spatial derivatives acting on them. A quick analysis of the geodesic equation reveals that the linear equations are indeed sufficient to accurately compute the photon transport. Note that this is no longer the case when using for example a comoving Nm gauge with $v=B$ which requires higher-order corrections in the weak-field limit. 
In addition we will, for the remainder of this work, neglect vector and tensor metric perturbations since they are generally sub-leading. However, these could be included using similar techniques.

The observables we compute are gauge invariant by definition. This means that we may work in the Nm gauge directly, or in any other gauge of our choice and link the perturbations in that gauge to the simulation. Since the expressions in the Poisson gauge are well known and relatively simple we will often use them to construct observables and then relate the simulation output to the Poisson gauge instead.

 Clearly, any observable $\mathfrak o$ may be written as
\be
\mathfrak o = \mathfrak{o}^{\rm P}( \Psi, \delta^{\rm P}, .... ) = \mathfrak{o}^{\rm P}( \Psi(A, \delta, ....), \delta^{\rm P}(A, \delta, ....), .... ) \,,
\ee
where $\mathfrak{o}^{\rm P}$ is the algebraic structure $\mathfrak o$ takes when expressed in terms of Poisson gauge perturbations, and in the rightmost equation we have expressed the Poisson gauge perturbations in terms of Nm gauge quantities employing a gauge transformation. These relations have been derived in our GR dictionary~\ref{sec:dic}.
In the simplest case, either in a pure dust Universe or when employing a backscaling Nm gauge at late times, this simplifies to
\begin{align} \label{eq:densdic}
\delta^{\rm P} &= \delta^{\rm N} - 3\Phi^{\rm N} +  3\zeta \,,\\ \label{eq:potdic}
\Phi &= \Phi^{\rm N} \,,\\
\Psi &= -\Phi =  -\Phi^{\rm N}\,, \\ \label{eq:vdic}
v^{\rm P} &= v^{\rm N} \,. 
\end{align}

\subsection{Magnification and shear}
In Poisson gauge magnification and shear are expressed in terms of the velocity and metric potentials alone.
For example the magnification can be written as~\cite{Challinor:2011bk,Jeong:2014ufa}:
\begin{align} \nonumber
\mathcal{M}^{\rm P} &= 1 -2\Phi + 2 \left(1 - \frac 1 {\Hc \chi_s}\right) \left(\Psi_e -\Psi_0 - v^{\rm P}_{||,e} + v^{\rm P}_{||,0} \right) \\ \nonumber
&- \frac 2 \chi_s \int \limits_{0}^{\chi_s} {\rm d}\chi (\Psi - \Phi) + \int \limits_{0}^{\chi_s} {\rm d}\chi (\chi_s - \chi)\frac \chi \chi_s \Delta_{\perp}(\Psi - \Phi)  \\
&+ 2 \left(1 - \frac 1 {\Hc \chi_s}\right) \int \limits_0^{\chi_s} {\rm d}\chi (\dot{\Psi}- \dot{\Phi} ) \,.
\end{align}
Here and in the following subscripts of $0$ and $e$ label the values at observation and emission respectively. The integration runs over a straight background geodesic from the position of the observer (here at $\fett{x} = 0$ for simplicity) to the observed position $\chi_s$. We have defined parallel and orthogonal quantities based on the direction of observation $n^i$ such that 
\be
v_{||} = n_i v^i + n_i \hat\nab^i v = n_i v^i -  \nab_{||} \kop^{-1} v   \,,
\ee
while the derivative $\nab^i_{\perp}$ is the projection of $\nab^i$ orthogonal to $n^i$.

Using the dictionary equations~(\ref{eq:Pphi})--(\ref{eq:Pv}) we obtain the magnification expressed using the simulation perturbations.
This relation is the unique connection between the simulation output and the gauge invariant magnification, valid in weak-field gravity from the large to the small scales. 
We find that the non-trivial metric of the Newtonian simulation induces corrections from radiation, appearing via the $\gamma$ term, that go beyond the information contained in the simulation potential $\Phi^{\rm N}$. These describe the impact of the non-trivial evolving metric on the light transport beyond the ``Newtonian'' bending of light by the simulation potential $\Phi^{\rm N}$.

However, in the special case of a back-scaled simulation at late times ($\gamma \approx 0$) we find that 
these corrections disappear. In particular we may employ the very simple relations~(\ref{eq:densdic}--\ref{eq:vdic}) stating that only the simulation density is not equivalent to the Poisson gauge density, while all other perturbations coincide ($\Psi = -\Phi =  -\Phi^{\rm N}$ and $v^{\rm P} = v^{\rm N}$). This implies that all observables $\mathfrak{o}^{\rm P}$ that do not depend explicitly on the density may be expressed using the unmodified Poisson gauge expressions together with the simulation output.

For this reason the magnification and shear in back-scaled simulations may, for sources at the late times, be computed from the simulation potential $\Phi^{\rm N}$ assuming a simple Poisson type metric. Note that this does not extend to the very early times, where $\HT$ evolves even when employing backscaling (see \cite{Fidler:2017ebh}). While negligible for galaxy lensing this might be relevant for high precision CMB lensing. 

\subsection{Ray tracing}

In contrast to the magnification and shear, which depend only on the metric perturbations, the observable number counts of tracers rely on the fully non-linear densities and the linear formula may not be used reliably. Alternatively, 
ray tracing can be employed to obtain the observable distribution of tracers on our celestial sphere. Based on the non-linear particle distribution in a numerical simulation, light rays are followed through the space-time and are used to construct the observables. The bending of these rays itself is based only on the small metric potentials and may be described accurately in a linear approximation based on the weak-field potentials. 

While for gauge independent observables we may employ a gauge of our choice and use the dictionary to relate the simulation perturbations with the relativistic perturbations, gauge dependent quantities must be computed in the gauge of the simulation, i.e. the Nm gauge. 
The position of a tracer $x^{\mu}$ is gauge dependent, but together with the gauge dependent light transport to the observer $\delta x^{\mu}$ we obtain the gauge independent position of the tracer on our celestial sphere characterised by the direction of observation $n^i$ and the comoving distance $\chi_s$ (or redshift).

Therefore, when computing the photon displacement $\delta x^\mu$ numerically from the simulation we may not perform the computation in Poisson gauge using the dictionary, but we must compute it directly in the Nm gauge. Alternatively we may gauge-transform the entire simulation output to the Poisson gauge (cf.\ \cite{Adamek:2017grt}), in which we obtain new positions for the galaxies and now the light transport is given by the well known Poisson gauge equations. 

To compute the leading photon displacement we integrate along a background geodesic parametrised by $\chi$ and obtain in Nm gauge 
\begin{align} 
\delta x^0 &= \left[-\delta a_0 - A_0 + v_{||}^0 +\frac{1}{2} \nab_{||}\kop^{-2}\dot{H}_{{\rm T}0}\right]\chi \nonumber \\
&+\int \limits_0^{\chi_s} {\rm d}\chi\Big[ 2 A + (\chi_s -\chi)(\dot{A} - \dot{\Phi} - \frac{1}{2} \nab_{||}^2\kop^{-2} \dot{H}_{\rm T})\Big] - \int_0^{\tau_0} {\rm d}\tau A(0,\tau) \,, \\ \nonumber
\delta x^i &= \left[\delta a_0 n^i +\Phi^0 n^i - \nab_{||}\hat\nab^i \kop^{-1} H_{{\rm T}0} - \frac{1}{2} \hat\nab^i \kop^{-1}\dot{H}_{{\rm T}0}- v_0^i - \hat\nab^i v_0\right]\chi\\ \nonumber
 &+\int_0^{\chi_s} {\rm d}\chi \bigg[ \frac{1}{2} \hat\nab^i \kop^{-1} \dot{H}_{\rm T} -2\Phi n^i + 2\nab_{||}\hat\nab^i \kop^{-1} \HT  \\
 & \phantom{+\int_0^{\chi_s} {\rm d}\chi \bigg[}+ (\chi_s -\chi) \nab^i \Big( \Phi -A +\frac{1}{2} \nab_{||} \kop^{-2} \dot{H}_{\rm T} + \nab_{||}^2 \kop^{-2}\HT  \Big) \bigg] \,,
\end{align}
where $\delta a_0$ is the difference of the scale factor at the observation from the background value, see \cite{Jeong:2014ufa}. 
Here we may relate the Nm gauge perturbations to the simulation potential $\Phi^{\rm N}$ using relations~(\ref{eq:NMA})--(\ref{eq:NMd}) in our GR dictionary.
We find that the presence of the metric potential $\HT$ causes an integrated coordinate shift along the photon path which is not present in a Poisson gauge analysis. 

Of particular interest is the more simple case of a simulation that makes use of backscaled initial conditions (cf.\ \cite{Fidler:2017ebh}).
In that case we have $\dot{H}_{\rm T} = 0$ which significantly simplifies the equations, but the correction from $\HT$ remains. For a time-independent field the derivative in the photon direction is also the derivative along the line-of-sight $\nab_{||} \HT = \frac {d}{d\chi} \HT$. This allows us to integrate the terms involving $\HT$ explicitly and we find
\begin{align}
\delta x^0 =& \left[-\delta a_0 + \Phi^{\rm N}_0 + v_{||}^0\right]\chi -2\int_0^{\chi_s} {\rm d}\chi\Big[\Phi^{\rm N} + (\chi_s -\chi)\dot{\Phi}^N\Big] + \int_0^{\tau_0} {\rm d}\tau \Phi^{\rm N}(0,\tau)\,, \\ \nonumber
\delta x^i =& \left[\delta a_0 n^i +\Phi^{\rm N}_0 n^i - v_0^i - \hat\nab^i v_0 \right]\chi +2\int_0^{\chi_s} {\rm d}\chi\Big[-\Phi^{\rm N} n^i + (\chi_s -\chi)\nab^i \Phi^{\rm N}\Big]\\ 
& + \hat\nab^i\kop^{-1}(H_{{\rm T}e}- H_{{\rm T} 0})\,. \label{LoSwithICS}
\end{align}
These equations, given in the coordinates of unmodified N-body simulations, 
describe the well-known effect stemming from the bending of light due to the scalar potential $\Phi^{\rm N}$, plus a novel term that arises through an Integrated Coordinate Shift (ICS) on the photon trajectories. The ICS, when using backscaled initial conditions, depends only on the difference of $\HT$ between the point of emission and the point of absorption (as a pure ``potential'' effect similar to the Sachs--Wolfe effect). A photon emitted at a given value of $\HT$ and later absorbed at a different value will be deflected along its line of sight as a consequence of the ICS effect. 

\begin{figure}[tb]
  \centering
    \includegraphics[width=0.6\textwidth]{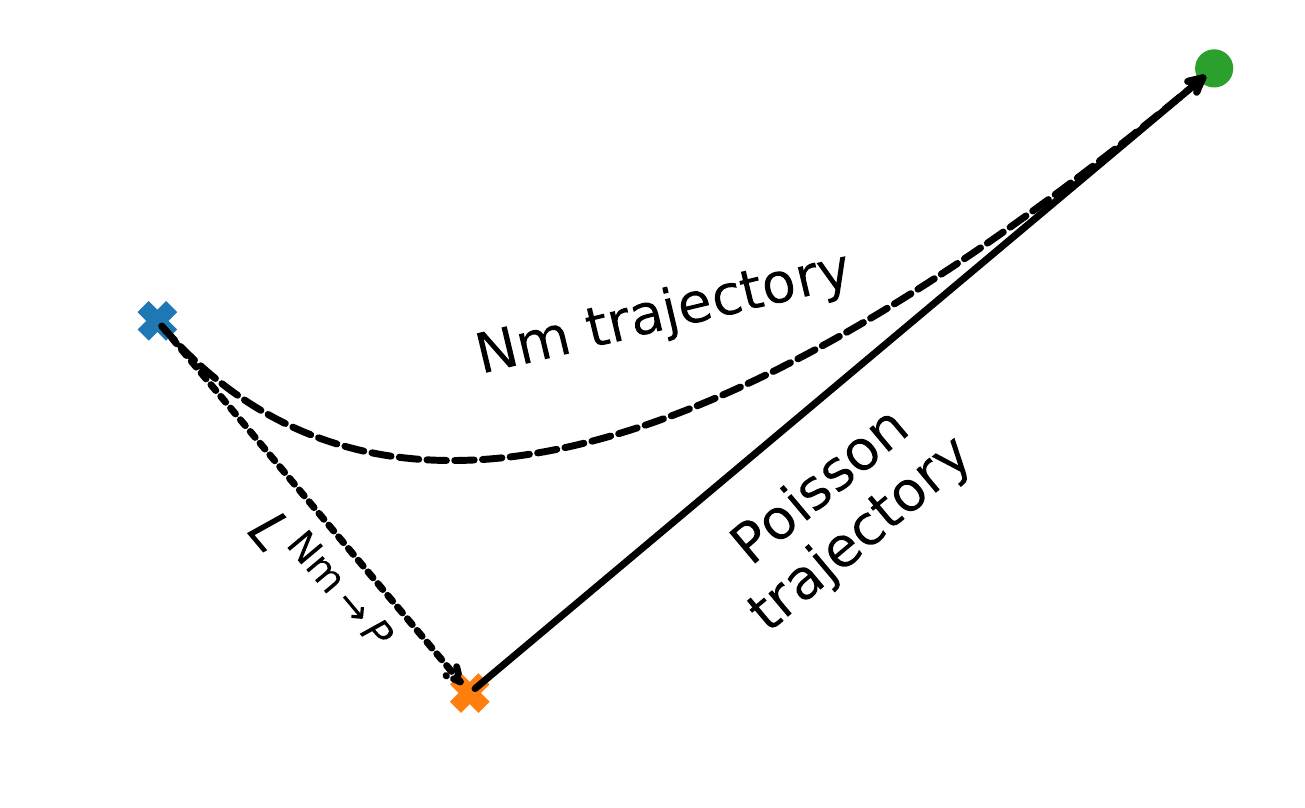}
  \caption{In the Nm gauge the ICS effect changes the direction of photons between emission and absorption, making them appear as originating from a different point of emission. In the Poisson gauge on the other hand, the spatial gauge transformation ($\fett{L}^{\rm Nm \to P} = \fett{x}^{\rm Nm} - \fett{x}^{\rm P}$) connecting both gauges puts the emitting particle exactly at that apparent position, while there is no coordinate shift. On the level of observation, combining the point of emission and the light propagation, both gauges agree. }
  \label{fig:disrupt}
\end{figure}

The impact of $\HT$ on photon geodesics is therefore indistinguishable from a change in the emitting matter density. As illustrated in Fig.~\ref{fig:disrupt}, if we transform the simulation output to the Poisson gauge, the ICS disappears since $\HT$ vanishes in the Poisson gauge. The spatial gauge transformation is given by $L^i = \kop^{-1} \hat\nab^i \HT$ and reshuffles the sources, exactly reproducing the signatures of the ICS effect. The impact of $\HT$ is now included in the emitting Poisson gauge density $\delta_\text{count}^{\rm P} =\delta^{\rm N} + \HT$. 
In the Nm gauge certain regions appear denser as a result of the ICS, while in Poisson gauge they \emph{are} denser due to the relative displacement from the spatial gauge transformation. On the level of the final observable both interpretations agree. 

In the previous section we have shown the importance of the non-trivial geometry from $\HT$ to the relativistic density. 
We now see that the same effect is also present in ray tracing where it appears as an integrated coordinate shift. In the same way that a Poisson gauge interpretation based only on $\delta^\text{N}$ and $\Phi^\text{N}$ provides incorrect results for the density on the large scales, a computation of observed number counts based on only $\Phi^\text{N}$ becomes incorrect. Ignoring the ICS is equivalent to approximating $\delta_\text{count}^{\rm P} \approx \delta^{\rm N}$ instead of 
\be
\delta_\text{count}^{\rm P} =\delta^{\rm N} + \HT =\delta^{\rm N} + 3 \zeta\,.
\label{eq:ICS}
\ee
The resulting error is comparable to that in the second line of figure~\ref{fig:dirupt}, employing the approximation $\delta^{\rm P} \approx \delta^{\rm N} - 3\Phi^{\rm N}$ instead of the much more accurate $\delta^{\rm P} \approx \delta^{\rm N} - 3\Phi^{\rm N} + \HT$.
In both cases, neglecting the metric potential $\HT$ induces a significant mismatch on the large scales, showing that the ICS is an important relativistic effect. 

Note that while gauge independent observables such as the magnification and shear discussed in the previous section are not affected by gravitational potentials beyond $\Phi^{\rm N}$, the lensing deflection discussed here is not an observable. Only together with the also gauge dependent galaxy positions we obtain a gauge independent observable. Therefore it is important to include the ICS when performing ray tracing directly on the galaxy positions extracted from Newtonian N-body simulations. 

\section{Comparison to related work in the literature}\label{sec:CZ}
In this section we discuss the connection between our work, the analytic method of~\cite{Chisari:2011iq}, the relativistic ray tracing code {\sc liger} and the relativistic N-body code {\it gevolution}.
\subsection{The analytic method of Chisari and Zaldarriaga}
In the previous work by Chisari and Zaldarriaga (CZ) \cite{Chisari:2011iq}, a dictionary between simulations and general relativistic perturbations in the Poisson gauge has been developed, using a different ansatz restricted to linear perturbation theory in a pressureless Universe. As expected, their findings correspond to our more general result in the appropriate limits.

In detail, they find that the potentials in the simulation are in agreement with the Bardeen potentials (see equation~\eqref{eq:potdic} in our framework), while the particles are not at the relativistic positions within the Poisson gauge. 
They argued that an initial displacement should be added to the coordinates of particles in a Newtonian simulation $\fett{x}^{\rm Nm}$ in order to obtain the coordinates in the Poison gauge 
\be
 \fett{x}^{\rm P} = \fett{x}^{\rm Nm} + \delta  \fett{x}_{\rm in}, \quad \nabla \cdot \delta  \fett{x}_{\rm in} = -5 \Phi^{\rm N}_{\rm in},
\ee
where $\Phi^{\rm N}_{\rm in}$ is the Newtonian potential at the initial time of the simulation. This initial displacement generates a density 
\be
\delta_{\rm in} = -\nabla \cdot \delta  \fett{x}_{\rm in} = 5 \Phi^N_{\rm in} = 3 \zeta, 
\ee
which should be added to the counting density in the Newtonian simulation to obtain the Poisson gauge density. 

While the methods are constructed differently, both approaches are mathematically equivalent (in the appropriate limits). The displacement suggested by CZ may be understood as the spatial gauge transformation connecting the Nm and Poisson gauge. While the relativistic density is identical in both gauges ($\delta^\text{Nm} = \delta^{\rm P}$ due to the shared temporal gauge), the simulation density $\delta^{\rm N}$ is affected by a spatial gauge transformation. The term $\HT$ appearing in our equation~\eqref{eq:ICS} 
exactly mimics the impact the displacement suggested by CZ would have on the coordinate density in the simulation.

We also find an agreement with our ray tracing computation. In the framework of CZ, there is no ICS effect, but the particles have to be displaced after the simulation is completed. In the last section we have shown that both, the ICS or an equivalent displacement of the particles, are identical in the limit considered in their work.

Beyond this agreement there are also differences. First, we are considering a more general cosmology where we allow for relativistic species. In addition our framework includes weak-field corrections which are neglected in the description of CZ. In particular, their displacement is introduced as a modification of the initial conditions which must be removed again once the simulation is finished. It is therefore evaluated at the initial time, based on the initial particle positions. Instead in our approach, we find that the metric in equation~\eqref{eq:densdic} is evaluated at the source position and time. 

Since the displacement is based on $\zeta$, which is constant in the late Universe, evaluating at the final time or initial time changes only the position at which the particle is found and $\zeta$ is evaluated. To linear order, the effect of the particle propagation may be neglected in combination with the perturbatively small curvature perturbation as a higher-order correction and both methods are in agreement. But to weak-field order, the motion of particles is no longer assumed to be small and the metric must be evaluated at the correct spatial position, which according to the weak-field Nm approach is the final one.

In the equations of CZ \cite{Chisari:2011iq}, the comoving curvature perturbation $\zeta$ is further replaced by the potential $\Phi$ using the relation 
\be
3\zeta = 5 \Phi \,.
\ee
This relation is only valid in the absence of radiation and/or dark energy. At the times where N-body simulations are usually initialised, this is of course 
fairly accurate, but the same relation may no longer be applied at the final time.

\subsection{The {\sc liger} code}
A recent paper \cite{Borzyszkowski:2017ayl} introduces the relativistic code {\sc liger}. The purpose of {\sc liger} is to construct mock galaxy catalogues on the 
past light cone using unmodified Newtonian N-body simulations, whilst incorporating general relativistic effects to linear order in perturbation theory. In particular, {\sc liger} includes linear GR effects from ray tracing, magnification and shear.
To achieve this, the code uses as an input the final particle distribution of a Newtonian N-body simulation, 
in combination with the GR dictionary of CZ~\cite{Chisari:2011iq} that reads, in our notation, $\delta^{\rm N} = \delta^{\rm synch}$, $\Phi^{\rm N} = \Phi$, and $v_i^{\rm N}= v_i^{\rm P}$, where 'synch' stands for synchronous gauge. 
The ray tracing employed in 
\cite{Borzyszkowski:2017ayl}
requires quantities to be given in the coordinates of the Poisson gauge (see their eqs.\,(9)--(10)). 
However since, as we have argued in this paper, unmodified N-body simulations implicitly make use of relativistic coordinates in the Newtonian motion gauge, the particles are not in Poisson gauge but in the Nm gauge. As a consequence, it appears that the current version of {\sc liger} misses the correction 
to the observed galaxy positions arising from the ICS in the NM gauge, i.e., the last line of our eq.\,\eqref{LoSwithICS}.
If instead the Poisson gauge ray tracing is used, then the galaxy positions must be shifted from NM to Poisson gauge, as illustrated in figure~\ref{fig:dirupt}.
Neglecting the ICS, by contrast, leads to mismatches in the observed galaxy positions.

We note that {\sc liger} explicitly neglects potential terms which could influence galaxy clustering on scales comparable to the Hubble radius, and the ICS term is another relativistic correction that becomes relevant on these scales.

\subsection{The {\it gevolution} code}
Another work closely related to ours is~\cite{Adamek:2015eda}, where the weak-field equations are solved numerically in the relativistic N-body code {\it gevolution}, working in the Poisson gauge. While a slightly different weak-field expansion is employed (see our discussion after eq.\,(\ref{eq:countdens})), the resulting corrections are small. On the level of the particle motion {\it gevolution} thus obtains a precision comparable with a Newtonian N-body simulation interpreted in our weak field Nm metric. 

Beyond the particle motion, {\it gevolution} also keeps the terms relevant for the computation of key relativistic signatures, such as the difference of the scalar potentials as well as vector and tensor modes, corresponding to our $\Sigma_{\rm GR}$. We neglect these as we are primarily concerned with the description of the relativistic dark matter evolution where they only appear as a sub-leading correction.  
However, we may still employ a Newtonian N-body code to compute these within the Nm gauge framework. After completing the N-body simulation and computing the radiation perturbations and the Nm gauge potential $\HT$ in a linear Boltzmann code, we may reconstruct $\Sigma$ and $\Sigma_{\rm GR}$, the leading terms responsible for the difference in the potentials or sourcing the vector and tensor modes of the metric. The dynamical equations~\eqref{eq:vec}--\eqref{eq:ten} can then be integrated independently, while the feedback from directly including these modes in the N-body simulation is a subleading correction in the weak-field limit. 

In~\cite{Adamek:2017grt} we have compared the (linear) Nm gauge framework with the {\it gevolution} code where we find an excellent agreement. 
We conclude that a weak-field Nm gauge analysis is equivalent in precision to a computation in {\it gevolution} for the dark matter evolution, while significant work still needs to be invested in the Nm gauge framework for obtaining the vector and tensor modes to the same precision as in {\it gevolution}.

\section{Conclusions}\label{sec:conclude}

We have shown that unmodified Newtonian N-body simulations can be interpreted within a weak-field limit of general relativity in the standard $\Lambda$CDM cosmology. 
The usual (Newtonian) analysis of N-body simulations implicitly relies on the assumption that the metric potentials remain small, although density perturbations may become large on small (sub-Hubble) scales.
Our work reproduces previous results \cite{Green:2010qy,Green:2011wc,Chisari:2011iq,Fidler:2015npa} in the absence of radiation and/or non-linear corrections. Relativistic species were studied in our previous works \cite{Fidler:2016tir,Fidler:2017ebh}, while we now identify additional non-linear corrections from weak-field gravity.  

Our framework supplements non-linear N-body results with outputs from linear Einstein-Boltzmann codes to construct a relativistic coordinate system within which to interpret an unmodified Newtonian simulation, and provides the tools to interpret the N-body particles at their relativistic positions. We explicitly constructed a GR dictionary which allows us to connect the simulation output to their relativistic counterparts, and even to relativistic quantities that cannot be encompassed within a Newtonian description.  

We employ an analytic approach to determine the photon transport for observables based on metric perturbations only, such as the magnification and shear, and a numerical ray-tracing method for observables that depend directly on the matter perturbations such as the observable number counts.
In both cases, we find that the off-diagonal scalar component in the spatial metric $\HT$, not present in the Newtonian simulation, enters in the computation and yields a coordinate shift along the photon trajectory. 

In the simplifying limit of a radiation-free universe or when employing backscaling at sufficiently late times, the analysis simplifies significantly. The magnification and lensing shear may be extracted based on the simulation's Newtonian potential alone assuming Poisson gauge formulas, because the contributions from the off-diagonal spatial metric $\HT$ cancel. 
This implies that Newtonian simulations may directly be applied to compute CMB lensing, as long as the lensing potential is dominated by the late-time potentials.
On the other hand, for ray tracing simulations, $\HT$ enters as a total derivative in the absence of radiation and can be integrated analytically. This provides an integrated coordinate shift to the particle positions. However, in the general case, the impact of $\HT$ needs to be included along the actual geodesics. 

We find that the contribution from $\HT$ is sub-dominant on small scales. 
However on the large scales, the impact from $\HT$ is of leading order and crucial for a consistent interpretation of a Newtonian simulation. In fact, taking all relativistic corrections that can be reconstructed from the simulation potential $\Phi^{\rm N}$ into account, but not from $\HT$, can lead to results which are worse than neglecting relativistic corrections altogether. 

We have focused in this work on a standard model cosmology, where we find a simple relation between a Newtonian non-linear system and weak field relativity, exploiting a number of cancelations and simple late-time limits. In an extended model beyond $\Lambda$CDM this may change and the interpretation of Newtonian simulations may become significantly more complex. We have studied, within a first-order Newtonian motion approach, a case of decaying dark matter~\cite{Fidler:2017ebh} generating a significant amount of radiation in the late Universe. In such cases we may still construct a Newtonian motion gauge, but the resulting relativistic to Newtonian dictionary and analysis may deviate significantly from the usual Newtonian approach.
 
\section*{Acknowledgements}

We thank Cyril Pitrou for support in the use of {\sc xpand} and
Daniele Bertacca, Jacob Brandbyge, Steen Hannestad, Cristiano Porciani and Christophe Ringeval for useful discussions. Prior to submission, Julian Adamek kindly shared his manuscript on a closely related topic.
CF is supported by the Wallonia-Brussels Federation grant ARC11/15-040 and the Belgian
Federal Office for Science, Technical \& Cultural Affairs through the Interuniversity Attraction
Pole P7/37.  
TT acknowledges support from the Villum Foundation.
The work of CR is supported by the DFG through the Transregional Research Center TRR33 ``The Dark Universe''. 
RC, KK and DW are supported by the STFC grant ST/N000668/1.  
The work of KK has received funding from the European Research Council (ERC) under the European Union's Horizon 2020 research and innovation programme (grant agreement 646702 ``CosTesGrav").

\bibliographystyle{JHEP}
\bibliography{references}

\end{document}